\documentclass[twocolumn,footinbib,preprintnumbers,floatfix,aps,prl,10pt]%
{revtex4-1}
\pdfoutput=1

\usepackage[hyperindex,breaklinks=true,hyperfootnotes=false,bookmarks=false]%
{hyperref}
\usepackage{graphicx}
\usepackage{amsmath}
\usepackage{xcolor}
\usepackage{amssymb}
\usepackage[utf8]{inputenc}

\newcommand{\eq}[1]{Eq.~\eqref{eq:#1}}
\newcommand{\eqs}[2]{Eqs.~\eqref{eq:#1} and \eqref{eq:#2}}

\newcommand{\rcites}[1]{Refs.~\cite{#1}}
\newcommand{\rcite}[1]{Ref.~\cite{#1}}

\newcommand{\Mae}[3]{\bigl\langle#1\bigr\rvert#2\bigr\rvert#3\bigr\rangle}

\newcommand{\bnslash}{\bar{n}\!\!\!\slash}
\newcommand{\bare}{\mathrm{bare}}
\newcommand{\bn}{\bar{n}}

\newcommand{\df}{\mathrm{d}}

\newcommand{\eps}{\epsilon}

\newcommand{\nn}{\nonumber}

\newcommand{\cL}{{\mathcal L}}

\newcommand{\cusp}{\mathrm{cusp}}

\newcommand{\msb}{{\overline{\rm MS}}}
\newcommand{\as}{\alpha_s}

\newcommand{\one}{{(1)}}
\newcommand{\two}{{(2)}}
\newcommand{\three}{{(3)}}

\allowdisplaybreaks[2]

\begin{document}


\preprint{MITP/18-031}

\title{The Three-loop Quark Jet Function}

\author{Robin Br\"user}
\email{brueser@uni-mainz.de}
\author{Ze Long Liu}
\email{liu@uni-mainz.de}
\author{Maximilian Stahlhofen}
\email{mastahlh@uni-mainz.de}

\affiliation{PRISMA Cluster of Excellence, Institute of Physics, Johannes 
Gutenberg University, 55128 Mainz, Germany \vspace{0.5ex}}


\begin{abstract}
We calculate the massless quark jet function to three-loop order.
The quark jet function is a universal ingredient in SCET factorization for many 
collider and decay processes with quark initiated final state jets.
Our three-loop result contributes to the resummation for observables 
probing the invariant mass of final state quark jets at N$^3$LL$^\prime$.
It represents the first complete three-loop result for a factorization 
ingredient describing collinear radiation. Furthermore it constitutes a major 
component of the $N$-jettiness subtraction/slicing method at N$^3$LO, which 
eventually may enable the calculation of fully-differential cross sections with 
a colorful final state at this order. 
\end{abstract}

\maketitle



\noindent \textbf{Introduction.} 
In QCD processes involving highly energetic partons factorization plays a 
crucial role.
Most importantly it provides a mean to disentangle perturbative physics from 
nonperturbative physics. Ideally the nonperturbative effects can thus
be absorbed into universal (process-independent) functions as e.g. the parton 
distribution functions. 
More generally, whenever there is a strong hierarchy of scales one can hope to 
establish a factorization formula at leading order in the small scale ratio(s) 
that separates the physics happening at the different scales. Besides a 
considerable simplification this usually  allows to resum 
large logarithms of the small scale ratio(s) to all orders in perturbation 
theory via renormalization group equations (RGEs) for the individual 
factorization ingredients.
Such factorization formulae are conveniently derived in soft-collinear 
effective theory (SCET)~\cite{Bauer:2000ew, Bauer:2000yr, Bauer:2001ct, 
Bauer:2001yt, Bauer:2002nz, Beneke:2002ph}.

In the following we will consider decay or scattering processes with final 
state jets and a large scale hierarchy between the jet invariant masses 
($\sim \tau$) and the total center of mass energy ($\sim Q$).
The cross section differential in a generic observable $\tau$ that constrains 
the jet invariant mass then schematically takes the factorized form
\begin{align}
 \frac{\df \sigma}{\df \tau} &= H(Q) \times [B_a \otimes B_b \otimes J_{i_1}
\otimes ... \otimes J_{i_N} \otimes S](\tau) 
\label{eq:fact}
\end{align}
at leading order in $\tau/Q$ and all orders in $\alpha_s$.
The $\otimes$~symbol denotes a convolution of the type
\begin{align}
A(\tau) \otimes B(\tau) \equiv \int\! \df \tau'\, 
A^i(\tau-\tau')\, B(\tau') \,.
\label{eq:conv}
\end{align}
For concreteness we assume here a process with two incoming ($a,b$) and $N$ 
outgoing partons involved in the hard interaction, which is described by the 
hard function $H(Q)$, as e.g. observed in proton-proton collisions at the LHC.
For $\tau \ll Q$ the initial state radiation is then collimated along the two 
incoming beam directions and the final state radiation is collimated along
$N$ different jet directions. Wide-angle soft emissions are taken into account 
by the soft function $S(\tau)$. The beam functions $B_i(\tau)$ and the jet 
functions $J_i(\tau)$ describe the effects of collinear radiation in the beam 
and final state jets, respectively.
The functions $S$, $B_i$, and $J_i$ are universal in the sense that 
they are independent of the details of the hard process (e.g. the 
colorless final state). 
The collinear functions $B_i$ and $J_i$ are furthermore equal for any 
observable that in the collinear limit effectively reduces to a measurement of 
the jet invariant mass ($\tau \to \sqrt{s}$).
A prime example for such an observable obeying factorization~%
\footnote{Neglecting potential factorization breaking effects due to Glauber 
modes~\cite{Gaunt:2014ska,Zeng:2015iba,Rothstein:2016bsq,Schwartz:2018obd} in 
hadron collisions, 
which are however absent up to three loops in fixed-order perturbation 
theory~\cite{Gaunt:2015pea}.}
as in \eq{fact} 
is the $N$-jettiness event shape~\cite{Stewart:2010tn} including the special 
cases beam thrust~\cite{Stewart:2009yx} ($0$-jettiness) and 
thrust~\cite{Farhi:1977sg} ($\sim$ $2$-jettiness in lepton collisions).

In this letter we focus on the jet function $J_q(s)$ for the case that 
the corresponding hard parton initiating the jet is a massless (anti-)quark 
($i=q$). The SCET (quark) jet function was introduced in \rcite{Bauer:2001yt}. 
It can be defined in terms of standard QCD fields as~\cite{Becher:2006qw}
\begin{align}
 J_q(s) &= \frac{1}{\pi N_c} \, \mathrm{Im}\biggl[ \frac{i}{\bn \cdot p} \int\! 
\df^d x \;e^{-i p\cdot x} \; \nn\\
&\qquad \times \Mae{0}{T \, \mathrm{Tr}\Bigl[ \frac{\bnslash}{4} 
W^\dagger(0)\psi(0)\overline{\psi}(x)W(x) \Bigr]}{0} \biggr]\,,
\label{eq:Jqdef}
\end{align}
where $T$ is the time-ordering operator, 
$n^\mu$ is the lightlike jet direction ($\bn \cdot n=2, n^2=\bn^2=0$), $p^\mu$ 
is the jet momentum ($s \equiv p^2$), the trace is over color ($N_c=3$) and 
spinor indices, and 
\begin{align}
W(x) &= P\, \exp \biggl[i\, g \int_{-\infty}^0 \!\!  \df s \, \bn \!\cdot\! 
A(x+s \bn)  \biggr]
\end{align}
denotes a ($n$-collinear) Wilson line.
This definition implies that we can use standard QCD Feynman rules in the 
calculation of the jet function as soft radiation has already been decoupled 
from the (leading order) collinear SCET Lagrangian by means of a field 
redefinition~\cite{Bauer:2001yt}.
At one loop the quark jet function ($i=q$) was computed in 
\rcites{Bauer:2003pi,Bosch:2004th} and the gluon jet function ($i=g$) in 
\rcite{Becher:2009th}. The two-loop results for $J_q(s)$ and $J_g(s)$  were 
obtained in \rcite{Becher:2006qw} and \rcite{Becher:2010pd}, respectively.

The two-loop quark jet function contributes to a number of important cross 
section predictions with resummation beyond NNLL accuracy, e.g. in 
DIS~\cite{Becher:2006mr},
for thrust~\cite{Becher:2008cf,Abbate:2010xh,Abbate:2012jh}, 
C-parameter~\cite{Hoang:2014wka,Hoang:2015hka}, and heavy jet 
mass~\cite{Chien:2010kc}.
To improve their precision to full N$^3$LL$^\prime$ level~%
\footnote{For thrust, C-parameter, and heavy jet mass at N$^3$LL$^\prime$ also 
the respective soft function correction is currently missing.
In the primed counting scheme the N$^x$LO boundary terms of the functions in the 
factorization formula are included for N$^x$LL$^\prime$ accuracy. For details 
and advantages of this scheme see, e.g., \rcite{Almeida:2014uva}.} 
(in the peak region) the three-loop correction to the quark jet function is 
required.
This is particularly desirable for the latter three observables as they are 
used for precise determinations of $\alpha_s$ from $e^+e^-$ data.
Another good motivation to calculate $J_q(s)$ at three 
loops is the perspective to extend the $N$-jettiness (infrared) 
subtraction/slicing method~\cite{Boughezal:2015dva,Gaunt:2015pea}, which has 
been applied successfully to several NNLO processes with final state 
jets~\cite{Gao:2012ja,Boughezal:2015dva,Boughezal:2015aha,Boughezal:2015ded,
Berger:2016oht}, to N$^3$LO.\\


\noindent \textbf{Calculation.} 
We work in general covariant gauge with gauge parameter $\xi$, where $\xi=0$ 
corresponds to Feynman gauge, and use dimensional regularization ($d=4-2\eps$).
We generate the three-loop Feynman diagrams for $J_q(s)$ with 
{\tt qgraf}~\cite{Nogueira:1991ex}. 
The output is then further processed using a custom code that does the 
color, Lorentz, and Dirac algebra.
Our code also performs partial fractioning of products of the 
eikonal Wilson line propagators following the strategy outlined in 
\rcite{Pak:2011xt}, 
and finally maps the resulting terms onto (scalar) integral topologies with 
twelve linearly independent propagators/numerators.

For the integrals in each topology we then perform an integration-by-parts (IBP) 
reduction to master integrals (MIs) with the public computer program {\tt 
FIRE}~\cite{Smirnov:2014hma} in combination with {\tt 
LiteRed}~\cite{Lee:2012cn,Lee:2013mka}.
Next, we identify MIs that are related across the different topologies 
by shifts of the loop momenta.
In this way we find 34 MIs in five topologies.
This set of MIs however turns out to be redundant.
Four extra (one-to-one) IBP relations among the MIs are revealed by the 
following tricks:

The first is to search for identities between integrals based on their Feynman 
parameter representation using the algorithm of \rcite{Pak:2011xt}, which is 
implemented in the {\tt FindRules} command of {\tt FIRE}.
In practice we apply {\tt FindRules} to a large list of (test) integrals. 
The output are a number of equalities among these integrals, which must hold 
after IBP reduction.
In our case this required two more independent relations among the 34 MI 
candidates. 

Even more relations are found via dimensional 
recurrence~\cite{Tarasov:1996br,Lee:2009dh,Lee:2010wea}.
Compact formulae that relate a generic $d$-dimensional integral to a linear 
combination of either ($d+2$)-dimensional or ($d-2$)-dimensional integrals with 
the same kind (but different powers) of propagators have been 
derived using Baikov's representation~\cite{Baikov:1996iu} of Feynman 
integrals in \rcites{Lee:2009dh,Lee:2010wea}.
They are implemented in the {\tt RaisingDDR} and {\tt 
LoweringDDR} commands of {\tt LiteRed}, respectively.
By applying {\tt RaisingDDR} we can thus directly express the 34 
$d$-dimensional MI candidates as linear combinations of integrals in $d+2$ 
dimensions.
We then IBP reduce the output, and lower the dimension 
of the resulting integrals back to $d$ using the {\tt LoweringDDR} command. 
Comparing the result to the original integrals after another IBP reduction 
yields four equalities among the $34$ MIs, including the two relations found 
with the {\tt FindRules} trick.

We have checked all four extra relations analytically to the required order in 
$\eps$.
We are thus left with 30 MIs, which have maximally eight quadratic and two 
linear (Wilson line) propagators.
Expressing the full three-loop amplitude of the quark jet function in terms of 
these MIs the dependence on the gauge parameter $\xi$ manifestly vanishes as 
expected. This provides a first cross check of our setup.

The next step is to compute the MIs to high enough order in $\eps$.
For this we adopt a strategy similar to the one described in 
\rcites{Panzer:2014gra,vonManteuffel:2014qoa,vonManteuffel:2015gxa}.
First, we note that the dependence of the MIs on the two 
external scalar products $s=p^2$ and $\bn \cdot p$ is completely fixed by 
scaling properties. From reparametrization 
invariance~\cite{Chay:2002vy,Manohar:2002fd} and dimensional counting in 
\eq{Jqdef} we even know that the full three-loop contribution equals 
$[(-s-i0)^{-1-3\eps} \times \mathrm{const.}]$ before taking the imaginary part. 
In order to determine the constant we can therefore safely set $s=-1$ and $\bn 
\cdot p=1$ in the evaluation of the MIs for convenience.

We then switch to a new basis of MIs that are 
quasi-finite~\cite{vonManteuffel:2014qoa} in some integer dimension, in our case 
$d=4$ or $d=6$.
By `quasi-finite' we mean integrals that are either convergent or their 
divergence can be factored out in a simple way. 
We allow e.g. divergences ($\propto 1/\eps^n$) contained in the prefactor of 
the corresponding Feynman parameter representation of the integral or generated 
as an overall prefactor by integrating over the Feynman 
parameters associated with the linear propagators. 
The latter integrations are always straightforward to carry out. 
We are thus left with up to eight convergent Feynman parameter integrals to be 
done.

A quasi-finite basis can be constructed as follows.
The basic idea is to increase the dimension of a given IR divergent integral to 
$d=n-2\eps$ ($n \in \mathbb{Z}$, $n>4$) in order to render it IR 
(quasi-)finite. For our MIs $n = 6$ turned out to be sufficient. One can then 
carefully increase 
the power of some propagators (by one) to decrease the degree of UV divergence 
without generating new IR singularities.%
\footnote{Bubble-type subintegrals can of course be conveniently integrated 
out in $d$ dimensions beforehand.}
Once a quasi-finite integral is found in this way it can be related to the 
original MI plus integrals with less propagators in $d=4-2\eps$ dimensions 
by dimensional recurrence and another IBP reduction. In 
some (exceptional) cases the quasi-finite integral does not reduce to the 
original MI and one has to try another quasi-finite candidate.
An algorithm that for a given integral automatically determines a desired 
number of proper quasi-finite integrals in shifted spacetime dimensions is 
implemented in the public program 
\mbox{\tt Reduze}~\cite{vonManteuffel:2012np}.

To perform the remaining convergent Feynman parameter integrals of the 
quasi-finite MIs we first expand the integrands to sufficiently high order in 
$\eps$. After that we integrate them using the {\tt HyperInt} 
package~\cite{Panzer:2014caa}. This code automatically evaluates linearly 
reducible convergent (Feynman) integrals in terms of multiple polylogarithms. 
With the outlined procedure we were able to compute all 30 MIs 
analytically to the required order in $\eps$.
We have checked all MI results numerically using the sector decomposition 
programs {\tt FIESTA}~\cite{Smirnov:2015mct} and {\tt 
pySecDec}~\cite{Borowka:2017idc}.
For many of the MIs we also have obtained analytic results with the Mellin 
Barnes technique~\cite{Smirnov:1999gc,Tausk:1999vh} employing the 
{\tt MB} package~\cite{Czakon:2005rk,Smirnov:2009up} as well as the PSLQ 
algorithm~\cite{PSLQ}.
We found perfect agreement in all cases.

To complete the calculation of the bare three-loop contribution to $J_q(s)$ we 
have to take the imaginary part according to \eq{Jqdef} and consistently expand 
in $\eps$.
To this end we use
\begin{align}
\mathrm{Im} \Bigl[ (-s-i0)^{-1-a \eps}  \Bigr]  &= 
 - \sin (\pi a \eps)  \, \theta(s)\,  s^{-1-a \eps}
\end{align}
and
\begin{align}
 \mu^{2 a \eps}\,\theta(s)\, s^{-1-a \eps} &= - \frac{\delta(s)}{a \eps}  + 
\sum_{n=0}^{\infty} \frac{(-a\eps)^n}{n!} 
\frac{1}{\mu^2} \cL_{n}\Bigl(\frac{s}{\mu^2}\Bigr)
\end{align}
with the usual plus distributions defined as
\begin{align} \label{eq:cLn}
\cL_n(x)
&= \biggl[ \frac{\theta(x) \ln^n \!x}{x}\biggr]_+ \!\!
= \lim_{\eps \to 0} \frac{\df}{\df x}\biggl[ \theta(x- \eps)\frac{\ln^{n+1} 
x}{n+1} \biggr]\,.
\end{align}
Convolutions among the $\cL_n(s)$ take the form
\begin{align} \label{eq:ExpLnLm}
(\cL_m \otimes \cL_n)(s)
= V_{-1}^{mn}\, \delta(s) + \sum_{k=0}^{m+n+1} V_k^{mn}\, \cL_k(s)\,.
\end{align}
A generic expression for $V_k^{mn}$ is given in \rcite{Ligeti:2008ac}.\\


\noindent \textbf{Result.} 
Bare and renormalized jet functions are related by ($i=q,g$)
\begin{align}
J_i^\bare(s)  &= Z_J^i(s,\mu) \otimes J_i(s,\mu)
\,.
\end{align}
Throughout this work we employ the $\msb$ renormalization scheme.
The RGE of the jet function reads
\begin{align}
\label{eq:RGE}
\mu \frac{\df}{\df \mu} J_i(s, \mu) &= 
\gamma_J^i(s,\mu) \otimes J_i(s, \mu) \,,
\end{align}
with the anomalous dimension
\begin{align} \label{eq:anomdim}
\gamma_J^i(s, \mu)
&= - (Z_J^i)^{-1}(s, \mu)  \otimes \mu \frac{\df}{\df \mu} Z_J^i(s,\mu)
\\
&= -2 \Gamma^i_{\cusp}(\as)\,\frac{1}{\mu^2}\cL_0\Bigl(\frac{s}{\mu^2}
\Bigr) + \gamma_J^i(\as)\,\delta(s)
\,.\end{align}
The collinear jet anomalous dimension $\gamma_J^i(\as)$ is equal to the one of 
the (virtuality-dependent) beam function~\cite{Stewart:2010qs}.
In the following we use the expansions
\begin{align}
\Gamma_\cusp^i(\alpha_s) &= \sum_{n=0}^\infty \Gamma^i_n 
\Bigl(\frac{\alpha_s}{4\pi}\Bigr)^{n+1}
,\;
\gamma^i_J(\alpha_s) = \sum_{n=0}^\infty \gamma_{n}^i 
\Bigl(\frac{\alpha_s}{4\pi}\Bigr)^{n+1},
\nn\\
J_i(s, \mu) &= \sum_{n=0}^\infty \Bigl(\frac{\alpha_s(\mu)}{4\pi} \Bigr)^n\, 
J_i^{(n)}(s, \mu)
\,.
\label{eq:Jiexp}
\end{align}
The coefficients of the (lightlike) cusp anomalous dimension 
($\Gamma^q_n$)~\cite{Korchemsky:1987wg,Moch:2004pa}%
\footnote{Due to Casimir scaling $\Gamma^g_n = C_A/C_F \, \Gamma^q_n$ for 
$n=0,1,2$. Beyond three loops Casimir scaling is violated, 
cf.~\rcites{Boels:2017skl,Moch:2017uml,Grozin:2017css}.}
and the collinear jet anomalous dimension ($\gamma_{n}^q$)~\cite{Becher:2006mr} 
for $n=0,1,2$ are e.g. listed in \rcite{Stewart:2010qs}.
The jet function coefficients have the form
\begin{align}
\label{eq:jetcoeffs}
J_i^{(m)}(s, \mu)
= J_{i,-1}^{(m)}\, \delta(s) + \sum_{n = 0}^{2m-1} J_{i,n}^{(m)}\, 
\frac{1}{\mu^2} \cL_{n}\Bigl(\frac{s}{\mu^2}\Bigr)
\,.
\end{align}
The coefficients of the $\mu$-dependent plus distributions in \eq{jetcoeffs} 
can be expressed in terms of lower-loop coefficients and anomalous dimensions 
by iteratively solving the RGE in \eq{RGE}. Up to three loops we find
\begin{align} 
J_{i,1}^\one &= \Gamma_0^i
\,,\nn\\
J_{i, 0}^\one &= - \frac{\gamma^i_0}{2}
\,,\nn\\
J_{i,3}^\two &= \frac{(\Gamma_0^i)^2}{2}
\,,\nn\\
J_{i,2}^\two &=
-\frac{\Gamma_0^i}{2} \Bigl(\frac{3\gamma^i_0}{2} + \beta_0 \Bigr)
\,,\nn\\
J_{i,1}^\two &=
\Gamma_1^i - (\Gamma_0^i)^2 \frac{\pi^2}{6}
+ \frac{\gamma^i_0}{2} \Bigl( \frac{\gamma^i_0}{2} + \beta_0\Bigr)
+ \Gamma_0^i\, J_{i,-1}^\one
\,,\nn\\
J_{i,0}^\two
&= (\Gamma_0^i)^2 \zeta_3 + \Gamma_0^i \gamma^i_0\, \frac{\pi^2}{12} - 
\frac{\gamma^i_1}{2}
- \Bigl(\frac{\gamma^i_0}{2} + \beta_0\Bigr) J_{i,-1}^\one 
\,,\nn\\
J_{i,5}^\three
&=\frac{(\Gamma_0^i)^3}{8}
\,,\nn\\
J_{i,4}^\three
&=-\frac{5}{12} (\Gamma_0^i)^2 (\frac{3}{4} \gamma_0^i + \beta_0)
\,,\nn\\
J_{i,3}^\three
&=\frac{\Gamma_0^i}{6}  \Bigl(5 \beta_0 \gamma_0^i +2 \beta_0^2 + \frac32 
(\gamma_0^i)^2 - \pi^2 (\Gamma_0^i)^2 + 6 \Gamma_1^i 
\nn\\
&\quad + 3 \Gamma_0^i 
J_{i,-1}^\one  \Bigr)
\,,\nn\\
J_{i,2}^\three
&=\frac{5}{2} (\Gamma_0^i)^3 \zeta_3 
+ \frac{\pi ^2}{4} (\Gamma_0^i)^2 (\gamma_0^i+\beta_0) 
- \frac{\Gamma_0^i}{4} (3 \gamma_1 + 2 \beta_1 )  
\nn\\
&\quad - \frac{\Gamma_1^i}{4} (3 \gamma_0^i + 4\beta_0) 
- \frac{\gamma_0^i}{16} \bigl(6 \beta_0 \gamma_0^i+8\beta_0^2
+ (\gamma_0^i)^2\bigr)
\nn\\
&\quad - \frac{\Gamma_0^i}{4} (3 \gamma_0^i + 8 \beta_0) J_{i,-1}^\one
\,,\nn\\
J_{i,1}^\three
&=\Gamma_2^i -\frac{\pi^4}{180} (\Gamma_0^i)^3 
-\zeta_3 (\Gamma_0^i)^2 (3 \beta_0 + 2 \gamma_0^i)
- \frac{\pi^2}{3} \Gamma_0^i \Gamma_1^i
\nn\\
&\quad - \frac{\pi^2}{12} \Gamma_0^i \gamma_0^i (3 \beta_0 + \gamma_0^i) 
+ \frac{\gamma_1^i}{2} (2\beta_0 + \gamma_0^i) + \frac{\beta_1 
\gamma_0^i}{2} 
\nn\\
&\quad +  
\Bigl(\frac32 \beta_0 \gamma_0^i+ 2 \beta_0^2 + \frac14 (\gamma_0^i)^2
-\frac{\pi^2}{6} (\Gamma_0^i)^2
+\Gamma_1^i \Bigr) J_{i,-1}^\one
\nn\\
&\quad +\Gamma_0^i J_{i,-1}^\two
\,,\nn\\
J_{i,0}^\three
&=\Bigl(3 \zeta_5-\frac{\pi ^2}{3}  \zeta_3 \Bigr) (\Gamma_0^i)^3 
+\frac{\pi^4}{90} (\Gamma_0^i)^2 \Bigl(\beta_0+ \frac{\gamma_0^i}{4} \Bigr)
\nn\\
&\quad
+\Gamma_1^i \Bigl(2 \zeta_3 \Gamma_0^i  +\frac{\pi^2}{12} \gamma_0^i \Bigr)
\nn\\
&\quad
+  \Gamma_0^i \Bigl( \frac{\pi^2}{12} \gamma_1 + \frac{\zeta_3}{2} \beta_0 
\gamma_0^i  
+ \frac{\zeta_3}{4} (\gamma_0^i)^2  \Bigr) - \frac{\gamma_2^i}{2}
\nn\\
&\quad +  \Bigl( \zeta_3 (\Gamma_0^i)^2 + \frac{\pi^2}{12} \Gamma_0^i 
(2\beta_0+  \gamma_0^i)-\beta_1 - \frac{\gamma_1}{2}  \Bigr) 
J_{i,-1}^\one
\nn\\
&\quad
- \frac12 (4\beta_0 + \gamma_0^i) J_{i,-1}^\two\
\label{eq:Jn}
\,,\end{align}
where $\beta_0 = \frac{11}{3}\,C_A -\frac{4}{3}\,T_F\,n_f$ and 
$\beta_1=\frac{34}{3}\,C_A^2  - (\frac{20}{3}\,C_A\, + 4C_F)T_F\,n_f$ with 
$n_f$ the number of active flavors.
Our result for the three-loop quark jet function perfectly reproduces 
\eq{Jn} for $i=q$. This provides another strong cross check and at the same 
time represents the first direct calculation of $\gamma_2^q$, 
which up to now has been inferred from RG consistency~\cite{Becher:2006mr} 
using the three-loop results of \rcites{Moch:2004pa,Moch:2005id}.
For $m=0,1,2$ the constants $J_{q,-1}^{(m)}$ are e.g. collected in 
\rcite{Gaunt:2015pea} in accordance with our conventions.
The new result of our work is
\begin{align}
 & J_{q,-1}^\three =
C_F^3 \Bigl(
274 \zeta_3
+\frac{22 \pi^2 \zeta_3}{3}
-\frac{400 \zeta_3^2}{3}
-88 \zeta_5
+\frac{1173}{8}
\nn\\&\qquad
-\frac{3505 \pi^2}{72}
+\frac{622 \pi^4}{45}
-\frac{9871 \pi^6}{8505}\Bigr)
\nn\\&
+C_A C_F^2 \Bigl(
-\frac{28241 \zeta_3}{27}
+\frac{2200 \pi^2 \zeta_3}{27}
+\frac{424 \zeta_3^2}{3}
+\frac{560 \zeta_5}{9}
\nn\\&\qquad
+\frac{206197}{324}
-\frac{17585 \pi^2}{72}
+\frac{18703 \pi^4}{1215}
+\frac{1547 \pi^6}{4860}\Bigr)
\nn\\&
+C_A^2 C_F \Bigl(
-\frac{187951 \zeta_3}{243}
+\frac{394 \pi^2 \zeta_3}{9}
+\frac{1528 \zeta_3^2}{9}
-\frac{380 \zeta_5}{9}
\nn\\&\qquad
+\frac{50602039}{52488}-\frac{464665 \pi^2}{4374}
+\frac{1009 \pi^4}{1620}
+\frac{221 \pi^6}{5103}\Bigr)
\nn\\&
+C_A C_F n_f T_F \Bigl(
\frac{14828 \zeta_3}{81}
-\frac{64 \pi^2 \zeta_3}{9}
+\frac{32 \zeta_5}{3}
-\frac{2942843}{6561}
\nn\\&\qquad
+\frac{136648 \pi^2}{2187}
-\frac{418 \pi^4}{405}\Bigr)
\nn\\&
+C_F^2 n_f T_F \Bigl(
\frac{22432 \zeta_3}{81}
-\frac{272 \pi^2 \zeta_3}{27}
+\frac{160 \zeta_5}{3}
-\frac{261587}{486}
\nn\\&\qquad
+\frac{4853 \pi^2}{54}
-\frac{5876 \pi^4}{1215}\Bigr)
\nn\\&
+C_F n_f^2 T_F^2 \Bigl(
\frac{1504 \zeta_3}{243}
+\frac{249806}{6561}
-\frac{1864 \pi^2}{243}
+\frac{8 \pi^4}{45}\Bigr)
\,.
\label{eq:Jqminusone3}
\end{align}
It is often convenient to work with the Laplace transform
\begin{align}
 \tilde J_q(\nu, \mu) &= \int_0^\infty \!\! \df s\,e^{-\nu s}  \, J_q(s)\,,
\end{align}
because the convolutions of \eq{conv} type turn in to simple 
products in Laplace space.
The Laplace space equivalents to our \eqs{jetcoeffs}{Jn} can 
be read off from \rcite{Becher:2008cf}.
The new three-loop constant related to \eq{Jqminusone3} 
in their  
notation is
\begin{align}
 &c^J_3 =
 25.06777873 \,C_F^3
 +32.81169125 \,C_A C_F^2
 \nn\\&\;
 -0.7795843561 \,C_A^2 C_F
 -31.65196210 \,C_A C_F n_f T_F
 \nn\\&\;
 -61.78995095 \,C_F^2 n_f T_F
 +28.49157341 \,C_F n_f^2 T_F^2\,,
\label{eq:cj3}
 \end{align}
where for the sake of brevity we have evaluated the exact analytical result to 
ten valid digits for each color factor.
The constant $c^J_3$ equals the position space coefficient $j_3$ 
affecting the $\alpha_s$ determinations in 
\rcites{Abbate:2010xh,Abbate:2012jh,Hoang:2015hka}, where until now $j_3=0 \pm 
3000$ has been assumed.
Evaluating \eq{cj3} for $N_c=3$, $T_F=1/2$ and $n_f=5$ we have
$j_3 = -128.6512525$.

In \rcite{Monni:2011gb} the N$^3$LO non-logarithmic constant of
the (normalized) thrust cumulant cross section in the singular limit was 
obtained from a fit to fixed-order data produced by the Monte Carlo program {\tt 
EERAD3}~\cite{GehrmannDeRidder:2007jk}, albeit with large numerical errors.
With our new three-loop jet function constant in \eq{Jqminusone3} and the known 
three-loop hard function~\cite{Abbate:2010xh} at hand we can use this result to 
extract a rough estimate for the unknown thrust ($q\bar q$ channel) soft 
function 
constant at 
three loops.
In Laplace (position) space and adopting the notation of \rcite{Becher:2008cf} 
($c^S_3$) and \rcite{Abbate:2010xh} ($s_3$) we find ($N_c=3$, $T_F=1/2$, 
$n_f=5$)
\begin{align}
c^S_3 \,= 2 s_3 + 691 = -19988 \pm 1440 \,(\text{stat.})  \pm 
4000\,(\text{syst.})
\,.
\end{align}


\noindent \textbf{Summary.} 
In this letter we have presented our calculation of the quark jet function 
$J_q(s)$ at three loops.
The main result is the three-loop contribution to the $\delta(s)$ coefficient 
and given in \eq{Jqminusone3}. All other terms at this order can be derived from 
RG consistency conditions in terms of previous results, see \eq{Jn}. 
The new contribution is a necessary ingredient to many N$^3$LL$^\prime$ 
resummed processes with final state jets. It has e.g. a direct impact on 
existing $\alpha_s$ determinations from $e^+ e^-$ event shapes.
Our calculation also represents the first step toward possible applications of 
the $N$-jettiness IR slicing (or subtraction)  method at N$^3$LO.

\bigskip
\begin{acknowledgments}
We thank Jonathan Gaunt for comments on the manuscript.
MS thanks Vicent Mateu and Robert Schabinger for helpful 
discussions.
RB is a recipient of a fellowship through GRK Symmetry Breaking (DFG/GRK 1581).
This work was supported in part by the Deutsche Forschungsgemeinschaft through 
the project ``Infrared and threshold effects in QCD'',
by a GFK fellowship and
the Cluster of Excellence Precision Physics, Fundamental Interactions and 
Structure of Matter (PRISMA – EXC 1098) at JGU Mainz.
\end{acknowledgments}

\bibliographystyle{apsrev4-1}
\bibliography{jetfunc}

\end{document}